\documentclass[%
aip,
 amsmath,amssymb,
 reprint,%
flushbottom,
]{revtex4-1}

\usepackage{graphicx}% Include figure files
\usepackage{dcolumn}% Align table columns on decimal point
\usepackage{bm}% bold math
\usepackage{booktabs}
\usepackage{siunitx}
\usepackage[utf8]{inputenc}
\usepackage[T1]{fontenc}
\usepackage{mathptmx}
\usepackage{etoolbox}
\usepackage{hyperref}
\hypersetup{
    colorlinks = true,
    linkcolor = blue,
    citecolor = blue,
    urlcolor=blue,
}

\makeatletter
\def\@email#1#2{%
 \endgroup
 \patchcmd{\titleblock@produce}
  {\frontmatter@RRAPformat}
  {\frontmatter@RRAPformat{\produce@RRAP{*#1\href{mailto:#2}{#2}}}\frontmatter@RRAPformat}
  {}{}
}%
\makeatother

\usepackage{setspace}
  \usepackage{parskip}
\usepackage[compact]{titlesec}         % you need this package
\titlespacing{\section}{6pt}{6pt}{6pt} % this reduces space between (sub)sections to 0pt, for example
\AtBeginDocument{%                     % this will reduce spaces between parts (above and below) of texts within a (sub)section to 0pt, for example - like between an 'eqnarray' and text
  \setlength\abovedisplayskip{4pt}
  \setlength\belowdisplayskip{4pt}}

\begin{document}

\preprint{AIP/123-QED}

\title{Time-resolved velocity and ion sound speed measurements from simultaneous bow shock imaging and inductive probe measurements}
% Force line breaks with \\
\author{R. Datta}
\affiliation{ Plasma Science \& Fusion Center, Massachusetts Institute of Technology, MA 02139, Cambridge, USA\looseness=-1 %\\This line break forced with \textbackslash\textbackslash
}%
 %\altaffiliation[Also at ]{Physics Department, XYZ University.}%Lines break automatically or can be forced with \\
\author{D. R. Russell}
\affiliation{ 
Blackett Laboratory, Imperial College London, London SW7 2BW, United Kingdom %\\This line break forced with \textbackslash\textbackslash
}%
\author{T. Clayson}
\affiliation{ 
Blackett Laboratory, Imperial College London, London SW7 2BW, United Kingdom %\\This line break forced with \textbackslash\textbackslash
}%
\author{J. P. Chittenden}
\affiliation{ 
Blackett Laboratory, Imperial College London, London SW7 2BW, United Kingdom %\\This line break forced with \textbackslash\textbackslash
}%
\author{S. V. Lebedev}
\affiliation{ 
Blackett Laboratory, Imperial College London, London SW7 2BW, United Kingdom %\\This line break forced with \textbackslash\textbackslash
}%
\author{J. D. Hare*}%
\email{jdhare@mit.edu.}
\affiliation{ Plasma Science \& Fusion Center, Massachusetts Institute of Technology, MA 02139, Cambridge, USA\looseness=-1 %\\This line break forced with \textbackslash\textbackslash
}%

%\author{C. Author}
% \homepage{http://www.Second.institution.edu/~Charlie.Author.}
%\affiliation{%
%Second institution and/or address%\\This line break forced% with \\
%}%

\date{\today}% It is always \today, today,
             %  but any date may be explicitly specified

\begin{abstract}
We present a technique to measure the time-resolved velocity and ion sound speed in magnetized, supersonic high-energy-density plasmas. We place an inductive (`b-dot') probe in a supersonic pulsed-power-driven plasma flow and measure the magnetic field advected by the plasma. As the magnetic Reynolds number is large ($R_M > 10$), the plasma flow advects a magnetic field proportional to the current at the load. This enables us to estimate the plasma flow velocity as a function of time from the delay between the current at the load and the signal at the probe. The supersonic flow also generates a detached hydrodynamic bow shock around the probe, the structure of which depends on the upstream sonic Mach number. By imaging the shock around the probe with a Mach-Zehnder interferometer, we determine the upstream Mach number from the shock Mach angle, which we then use to determine the ion sound speed from the known upstream velocity. We use the measured sound speed to infer the value of $\bar{Z}T_e$, where $\bar{Z}$ is the average ionization, and $T_e$ is the electron temperature. We use this diagnostic to measure the time-resolved velocity and sound speed of a supersonic $(M_S \sim 8)$, super-Alfvénic $(M_A \sim 2)$ aluminum plasma generated during the ablation stage of an exploding wire array on the MAGPIE generator (1.4 MA, 250 ns). Velocity and $\bar{Z}T_e$ measured using this technique agree well with optical Thompson scattering measurements reported in literature, and with 3D resistive MHD simulations in GORGON.

\end{abstract}

\maketitle

\newcommand{\ra}[1]{\renewcommand{\arraystretch}{#1}}

\setlength{\textfloatsep}{6pt plus 0.5pt minus 0.5pt}

\section{\label{sec:intro} Introduction}
To adequately diagnose laboratory plasma experiments, it is necessary to measure several plasma quantities within the finite lifetime of the generated plasma. As this often requires an extensive diagnostic suite, it can be advantageous if a single diagnostic can measure multiple quantities simultaneously.

In highly conducting plasmas with frozen-in magnetic flux, localized time-resolved measurements of magnetic field can also provide information about the plasma velocity. The inductive (`b-dot') probe is one of the simplest magnetic diagnostics --- a loop of wire that outputs a voltage due to time-varying magnetic flux. It provides a cheap and effective alternative to more advanced magnetic diagnostic techniques, such as Faraday rotation \cite{Hare2018} and proton radiography. \cite{Kugland2012,Sutcliffe2021}  B-dot probes have been used extensively to provide magnetic field measurements in magnetized high-energy-density (HED) plasmas. In laser-driven plasmas, which exhibit rapidly-evolving Biermann-generated or externally-imposed fields, multi-axis probes are used to reconstruct spatially- and temporally-resolved 3D magnetic fields. \cite{Everson2009,Pilgram2021, Levesque2022, Schaeffer2022}  Similarly, b-dot probes are routinely fielded in pulsed-power-driven plasmas to measure the $\sim 1-10$ T fields advected by the plasma flows. \cite{Burdiak2017,Suttle_2019}

The magnetic Reynolds number $R_M = UL/\bar{\eta}$ represents the relative importance of magnetic field advection over resistive diffusion. Here, $U$ is the plasma velocity, $L$ is the characteristic length scale, and $\bar{\eta}$ is the magnetic diffusivity. For large Reynolds numbers $R_M \gg 1$, advection dominates and the magnetic field becomes frozen into the flow. In magnetized plasmas with frozen-in flux, inductive probes can provide an estimate of the bulk flow velocity\cite{Ji1999,Pilgram2021} --- a quantity which is typically measured using other techniques, such as optical Thompson Scattering (OTS), \cite{Froula2006,Rosenberg2012,Suttle2021} Mach probes, \cite{Hutchinson2002} or from the Doppler shift of spectral lines.\cite{Suckewer1979} Inductive probe measurements have previously been used to estimate the expansion velocity of laser-driven $R_M \sim 10^4$ plasma bubbles, \cite{Pilgram2021} as well as inflow and outflow velocities in subsonic magnetic reconnection experiments. \cite{Yamada1997,Ji1999}

Flow velocity can be inferred by placing multiple probes along a velocity pathline, and measuring the transit time of a magnetized fluid parcel between the probes. However, in situations where the first probe significantly perturbs the velocity or magnetic field seen by the second probe, it may not be possible to position multiple probes along the same pathline. There are several ways to overcome this issue --- (1) a single probe can be used, and its location can be varied between repeated shots, \cite{Pilgram2021} (2) in symmetric systems, probes can be positioned on different pathlines along the direction of symmetry, and (3) the magnetic field can be measured at the source, such as in the transmission line of a pulsed-power machine, and a probe can be placed in the flow to measure the advected field.

\begin{figure}
\includegraphics[width=0.5\textwidth,page=1]{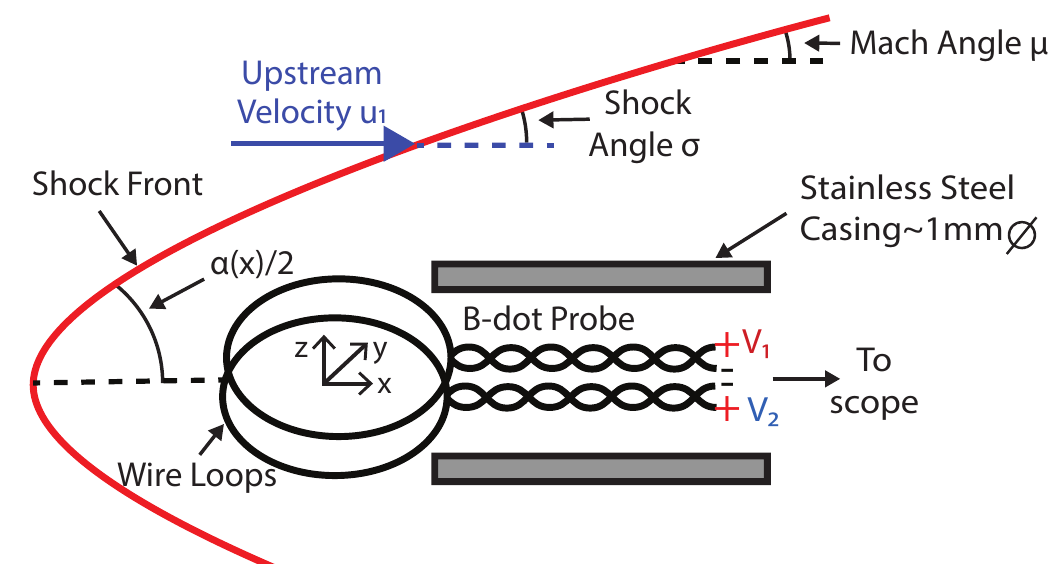}
\centering
\caption{B-dot probe with two oppositely-wound loops of enamel-coated copper wire, threaded through a thin-walled steel tube. The bow shock around the probe is represented by the red solid line.}
\label{fig:diagnostic_schematic}
\end{figure}

In addition to the flow velocity, inductive probes can also be used to estimate the Mach number and ion sound speed in supersonic plasmas. In a supersonic flow, the probe serves as an obstacle that generates a 3D bow shock, as illustrated in Figure \ref{fig:diagnostic_schematic}a. Bow shocks are curved discontinuities across which fluid parameters --- density, pressure, and velocity --- change abruptly. In magnetohydrodynamic (MHD) fluids, the magnetic field may also change discontinuously across the shock front. \cite{goedbloed_keppens_poedts_2010} The geometry of a bow shock depends on the upstream Mach number of the flow. \cite{Spreiter1969,anderson_2001,Burdiak2017} In hydrodynamic shocks, the relevant Mach number is the sonic Mach number $M_S = U / C_S$; in MHD shocks, however, the shock geometry depends on the both the sonic and the Alfvén Mach numbers, as well as on the orientation of the magnetic field. \cite{Spreiter1969,goedbloed_keppens_poedts_2010} Here, $C_S = \sqrt{\gamma \bar{Z} T_e/m_i}$ is the ion sound speed, which is a function of the adiabatic index of the plasma $\gamma$, ionization $\bar{Z}$, and the electron temperature $T_e$, and ion mass $m_i$. The Alfvén Mach number $M_A = U / V_A$ is the ratio of the flow velocity to the Alfvén speed $V_{\text{A}} = B/\sqrt{\mu_0 \rho}$, which depends on the magnetic field strength $B$ and density $\rho$ 

At the resistive diffusion length $l_\eta = \bar{\eta}/U$, the magnetic Reynolds number becomes $R_M = 1$, i.e. the rates of advection and diffusion become comparable. Therefore, if the obstacle size is smaller than or comparable to resistive diffusion length, the plasma and the magnetic field decouple at the shock. For inductive probes with diameters comparable to $l_\eta$, shocks around them will, therefore, be almost hydrodynamic, and jumps in the magnetic field will be negligible. By imaging the shock around a small inductive probe, the upstream sonic Mach number $M_S$ can be inferred, which allows us to further estimate the ion sound speed $C_S$ if the flow velocity is known. 

In this paper, we estimate the temporally-resolved velocity and the ion sound speed in a pulsed-power-driven magnetized HED plasma from simultaneous voltage measurements and imaging of bow shocks around small inductive probes. In \S \ref{sec:velocity_probe}, we outline the methodology of velocity estimation from b-dot measurements.  In \S \ref{sec:probe_design}, we describe the diagnostic setup fielded on the MAGPIE pulsed-power machine. In \S \ref{sec:velocity} and \S \ref{sec:imaging}, we present velocity and sound speed measurements made using this diagnostic. Finally, in \S \ref{sec:sims_lit}, we compare our measurements with 3D resistive MHD simulations, and with OTS measurements reported in literature.
\section{Velocity Measurement with B-dot Probes}
\label{sec:velocity_probe}
Consider a fluid parcel $x_p(x_0,t)$ which is at some initial position $x_0$ at time $t = 0$ in a magnetized high-$R_M$ plasma with a 1D time-varying velocity field $u(x,t)$ . The fluid parcel travels to a position $x_p(x_0,\tau) = x_0 + s$ after some time $\tau$. Using probes positioned at $x = x_0$ and $x = x_0 + s$, we measure the transit time $\tau$ of the fluid parcel. The average speed $\bar{u}$ of the fluid parcel along its trajectory over the time period $\tau$ is
\begin{equation}
    \bar{u} = \frac{s}{\tau} = \frac{1}{\tau}  \int_0^\tau \frac{dx_p(x_0,t)}{dt} dt
    \label{eq:speed}
\end{equation}

The expression within the integral is the Lagrangian velocity $\dot{x}_p(x_0,t)$ of the fluid parcel, which can be related to the Eulerian velocity field $u(x,t)$ using 
\begin{equation}
    \frac{dx_p}{dt} = u(x,t)|_{x = x_p(x_0,t)}
    \label{eq:FLE}
\end{equation}

 For this calculation, we assume a simple Eulerian velocity field $u(x,t) = (U_0 + mx) e^{\nu t}$. We choose this velocity field as it allows for a simple analytical result, and is a good approximation to the velocity field we simulate in the experiments described below, but the analysis outlined here can be performed for any arbitrary field. Substituting the given velocity field into Equation \ref{eq:FLE} results in a first-order non-linear ordinary differential equation, the solution to which is the trajectory of the fluid parcel in the prescribed velocity field.
\begin{equation}
    x_p(t) = -\frac{U_0}{m} + \frac{U_0 + mx_0}{m} \exp\left[\frac{m}{\nu}(e^{\nu t}-1)\right]
\end{equation}

With the known trajectory, we can calculate the average speed of the parcel from  Equation \ref{eq:speed}.
\begin{equation}
    \bar{u} = \frac{U_0 + mx_0}{m \tau}\left[ \exp\left\{\frac{m}{\nu}(e^{\nu t}-1)\right\}-1\right]
    \label{eq:average_velocity}
\end{equation}

The is the quantity measured from probe signals. In most applications, however, we are more interested in the instantaneous flow velocity, which is the Eulerian velocity. We can compare the average Lagrangian velocity described by Equation \ref{eq:average_velocity} to the flow velocity at the probe location $x = x_0 + s$, where the Eulerian flow velocity is $u_s \equiv u(x_p(\tau),\tau)$. Taking the ratio $\bar{u}/u_s$, we can calculate the extent to which the averaged Lagrangian velocity over- or underestimates the Eulerian velocity at the probe location. Furthermore, we are free to choose the position of the origin in this problem, so we set $x_0 = 0$, and the ratio $\bar{u}/u_s$ then simply becomes a function of the dimensionless quantities $m \tau$ and $\nu \tau$.
\begin{equation}
    R(m\tau, \nu \tau) \equiv \frac{\bar{u}}{u_s} = \frac{e^{-\nu \tau}}{m \tau}\left[1-\exp\left\{\frac{m \tau}{\nu \tau}\left(1-e^{\nu \tau}\right)\right\}\right]
    \label{eq:ratio}
\end{equation}

The value of $R$ depends on the spatial and temporal variation of the velocity field $u(x,t)$ between the points $x_0 < x < (x_0 + s)$ over the transit time $\tau$. The quantity $\bar{u}$ inferred from the probe measurements over-estimates the flow velocity at the probe location, i.e. $R > 1$, when the velocity field decreases both spatially ($m\tau < 0)$ and temporally ($\nu \tau < 0)$, while it under-estimates the flow velocity when the velocity field increases both spatially ($m\tau > 0)$ and temporally ($\nu \tau > 0)$. When $\nu$ and $m$ have opposite signs, the value of $R$ will depend on the relative magnitude of these terms. In the limit where both $\nu \tau \rightarrow 0$ and $m \tau \rightarrow 0$, $R = 1$, which shows that when the velocity field is spatially and temporally constant, or when the transit time $\tau$ is small, $\bar{u}$ becomes equal to the Eulerian velocity at the probe location. This means that reducing the separation of the probes will allow for a better estimate of the Eulerian velocity. Estimating the quantity $R$ requires some prior knowledge of the variation of the velocity field $u(x,t)$, which may be informed from simulations, previous experiments, or from analytical models. 

\section{\label{sec:probe_design} Diagnostic setup and demonstration on the MAGPIE pulsed-power machine}
\subsection{Inductive Probe Design}
The inductive probe used in this demonstration consists of two counter-wound single-turn loops of enamel coated copper wire, threaded through a $\sim \SI{1}{\milli \metre}$ OD thin-walled steel tube (Figure \ref{fig:diagnostic_schematic}). The two loops have diameters of $\sim \SI{0.5}{\milli \metre}$, and are coated with silver paint to shield them from electrostatic noise. The voltage signal from each loop is attenuated using a  voltage divider, and recorded separately using an oscilloscope with a 1 ns digitization rate.

The voltage $V_{1,2}$ across each loop can have two distinct contributions --- one due to the time-varying magnetic flux $\dot{\Phi}$ through the loop, and the other due to stray voltages $V_S$ from the pulsed-power generator: $V_{1,2} = \pm \dot{\Phi} + V_S$. Having two counter-wound loops allows us to combine the two signals and isolate the contribution of the time-varying magnetic flux, which is the product of the rate of change of the magnetic field  $\dot{B}$, and the effective loop area $A_{\text{eff}}$.
\begin{equation}
    V = 0.5(V_1 - V_2) = \dot{B} A_{\text{eff}}
\end{equation}

\begin{figure}
\includegraphics[width=0.48\textwidth,page=3]{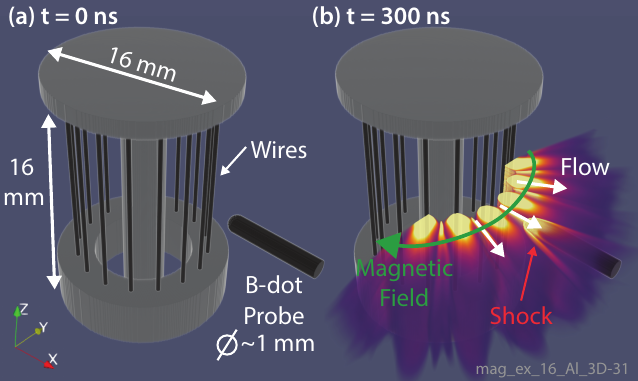}
\centering
\caption{(a) 3D representation of the experimental geometry, showing an exploding wire array with 16 equally-spaced \SI{30}{\micro\metre} Al wires (b) Ablation of plasma from the wires generates radially diverging flow which advects an azimuthal magnetic field. Left and top half of plasma has been clipped for clarity.}
\label{fig:experimental_setup}
\end{figure}

\subsection{Use on the MAGPIE Pulsed-Power Machine}

 Pulsed-power machines compress currents in space and time to deliver a large current pulse ($\sim 1-30$ MA) over $\sim 100-\SI{300}{\nano \second}$ time scales. A commonly-used load to generate radially-diverging magnetized outflows is the exploding wire array, which is a cylindrical cage of thin wires ($\sim 10-500$ \SI{}{\micro \metre} diameter) around a central cathode. When current flows through the wires, intense resistive heating converts the wires into hot dense ($T_i \sim T_e \sim 10$ eV, $n_e \sim 10^{18}$ \SI{}{\per \centi \metre \cubed}) plasma. The global magnetic field is oriented azimuthally and acts on the inner surface of the wires. The ${\bf j \times B}$ force accelerates the plasma radially outwards, creating radially-diverging flows of collisional ($\lambda_\text{mfp} / L \ll 1$) plasma. \cite{Lebedev2014, Burdiak2017, Hare2018, Suttle_2019, Russell2022} The magnetic Reynolds number (computed at the plasma length scale $\sim \SI{1}{\centi \meter}$) is typically large ($R_M > 10$), \cite{Burdiak2017,Suttle_2019} so the ablating plasma advects magnetic field from inside the array to the outside, creating flows with frozen-in azimuthally-oriented field lines. 

We position the probe to measure the azimuthal magnetic field advected by the plasma generated during the ablation stage of a 16 mm tall, 16 mm wide exploding wire array with 16 equally-spaced \SI{30}{\micro \metre} diameter aluminum wires (California Wire Company), as illustrated in Figure \ref{fig:experimental_setup}. The probe is positioned $5.55 \pm \SI{0.25}{\milli \metre}$ from the wires, with the loop normal along the magnetic field (i.e. in the y-direction). We use a coordinate system centered at the intersection of the obstacle axis and the array surface, with the magnetic field contained in the x-y plane. The MAGPIE generator (Imperial College London) drives a 1.4 MA peak current, 250 ns rise time current pulse through the load.\cite{Mitchell1996} Ablation of plasma from the wires generates hypersonic ($M_S \sim 8$) and super-Alfvénic ($M_A \sim 2$) outflows with frozen-in magnetic flux ($R_M \sim 20$).\cite{Burdiak2017,Suttle_2019} The resistive diffusion length is $l_\eta \sim 0.3 - 0.6$ mm, comparable to the probe diameter. We calculate $l_\eta$ using characteristic values of $U \sim \SI{60}{\kilo \metre \per \second}$, $n_e \sim10^{18} \SI{}{\per \cubic \centi \meter}$, and $T_e \sim 10 - 15$ eV. \cite{Suttle_2019}

In addition to the probe in the flow, we place a second single-loop b-dot probe in a recess inside MAGPIE's magnetically insulated transmission line (MITL). This probe monitors the current delivered to the load, which in turn sets the driving magnetic field, a fraction of which is advected by the flow. By measuring the time of flight between the driving and advected magnetic fields, we can estimate the flow velocity, as discussed in \S \ref{eq:average_velocity}.

\subsection{Bow Shock Imaging}

To image the shock around the probe, we use a Mach-Zehnder interferometer with a 1053 nm Nd:Glass laser (\SI{1}{\nano \second}, \SI{100}{\milli \joule}). The interferometer is set up to provide a line-integrated (along the y-direction) view of the side-on (x-z) plane, with a $\sim \SI{2}{\centi \metre}$ diameter field-of-view. We combine the probe and reference beams at the CCD of a Canon EOS 500D DSLR camera. When the probe beam propagates through the plasma, the resulting phase accumulated by the beam distorts the fringe pattern on the interferogram, and introduces a spatially-varying fringe shift,  which we use to reconstruct the phase difference between the probe and reference beams, and to calculate the line-integrated electron density. \citep{Swadling2013,Hare2019}

We use the side-on line of sight to eliminate the effect of magnetic field pile-up on the shock structure. Magnetic field pile-up occurs when advected field lines accumulate and bend around an obstacle. Magnetic tension of the bent field lines opposes the ram pressure of the upstream flow, creating a wider shock than when there is no pile-up. \cite{Dursi2008,Burdiak2017} For 3D obstacles like the inductive probe, pile-up  increases the opening angle of the shock in the plane parallel to the magnetic field (the x-y plane), but not in the plane normal to the field (the side-on x-z plane). \cite{Burdiak2017} This allows us to directly estimate the upstream Mach number from the side-on shock geometry. 

\subsection{3D Resistive MHD Simulations}

We perform 3D resistive MHD simulations of the experimental geometry using GORGON to compare velocity and sound speed measurements with simulated values. GORGON is a 3D (cartesian, cylindrical or polar coordinate) Eulerian two-temperature resistive MHD code with van Leer advection. \citep{Chittenden2004} GORGON can implement different radiation loss and ionization models. In this paper, we use a simple volumetric radiation-recombination model \cite{Huba2013} with a constant multiplier to account for line radiation, and an LTE Thomas-Fermi equation-of-state to determine the ionization level. We simulate an exploding wire array with the same geometric dimensions, wire material, and wire thickness, as in the experimental setup.  The simulation domain is a cuboid with dimensions $60 \times 60 \times $ \SI{45}{\milli\metre\cubed} and a resolution of $\SI{100}{\micro\metre}$. The initial mass in the wires is distributed over $\SI{300}{\micro \metre}$ wide pre-expanded wire cores. The applied current pulse (1.4 MA, 250 ns) was determined by integrating the Rogowski signal monitoring the current in the cathode for a shot with a similar load.

\begin{figure}
\includegraphics[width=0.48\textwidth,page=2]{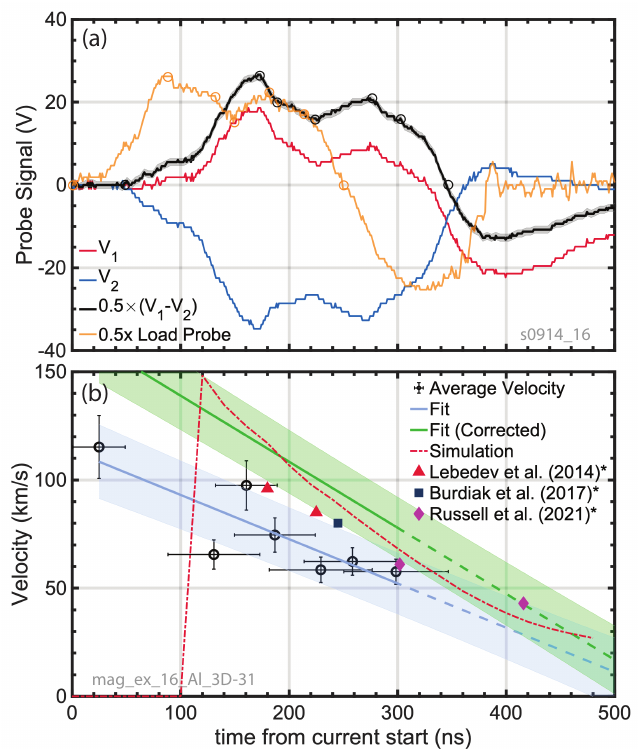}
\centering
\caption{(a) Voltage response of the inductive probe and load probe.
(b) Time-resolved average velocity measurements between the wires and probe location. Time error bars show period over which velocity is averaged. Data marked with an asterisk represent OTS data. The green curve represents corrected velocity using R = 0.67}
\label{fig:velocity}
\end{figure}

\begin{figure*}
\includegraphics[width=0.90\textwidth,page=4]{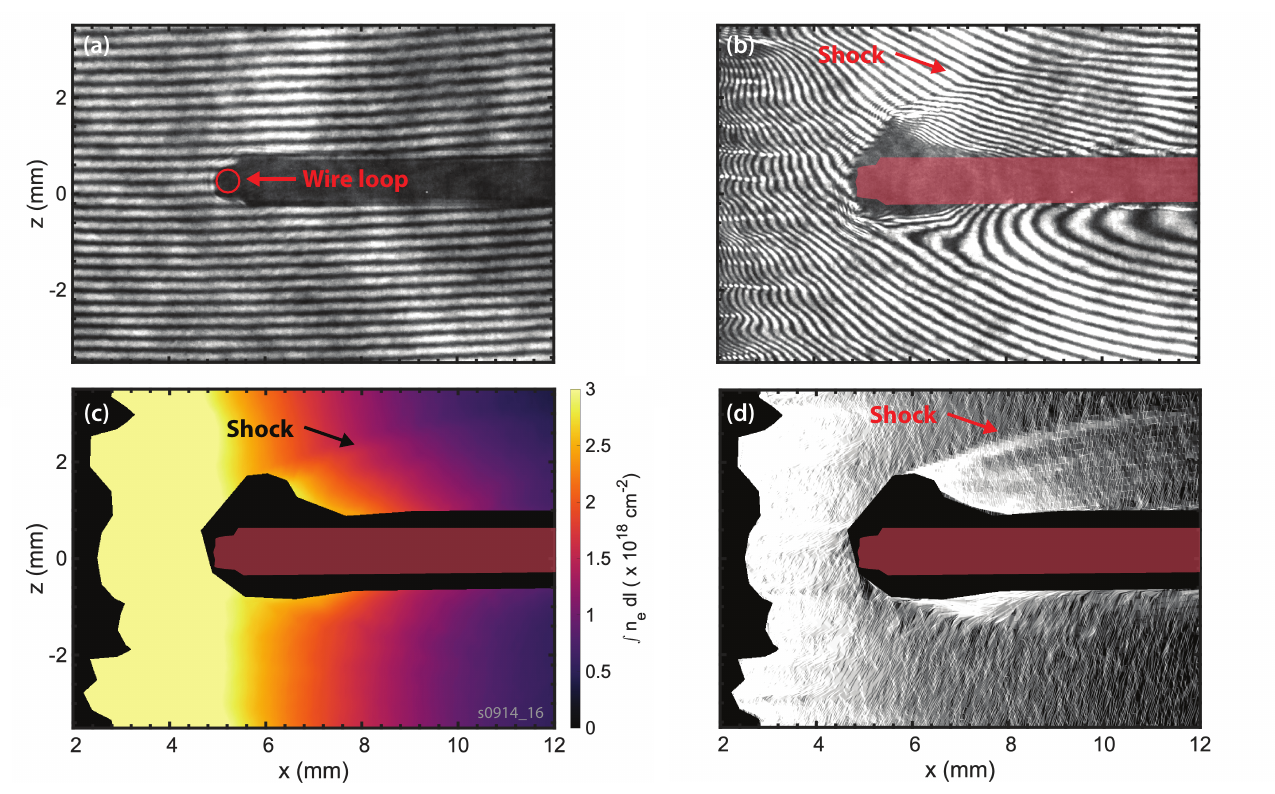}
\centering
\caption{\centering (a) Side-on background interferogram recorded before current start. (b) Side-on raw interferogram at 300 ns after current start. The red shaded region represents the silhouette of the probe from the background interferogram. (c) Side-on line-integrated electron density map. (d) Enhanced image of the shock. Fine-scale structures seen here are artifacts of interpolation. 
}
\label{fig:interferometry}
\end{figure*}

\section{\label{sec:velocity} Time-resolved Velocity Measurements}

Figure \ref{fig:velocity}b shows the unattenuated voltage signals from the oppositely-wound loops of the b-dot probe in the flow, together with the signal from the probe placed near the MITL (henceforth, referred to as the load probe).  We combine the oppositely-polarized signals from both loops to isolate the inductive signal of the probe (black curve in Figure \ref{fig:diagnostic_schematic}b). The probe in the flow reproduces the shape and characteristic features of the signal at the load, showing that the magnetic field is frozen into the flow.

The features present in the signals (represented by the orange and black circles on the load and flow signals respectively) are caused by voltage reflections due to impedance mismatches in the transmission lines of the pulsed-power machine. \cite{Mitchell1996} These features are typically considered undesirable, as they represent inefficient signal transmission. However, the presence of these features allows us to measure time-of-flight of the plasma to the probe from the time delay between corresponding features on the two signals. From the known distance between the array and the flow probe, we estimate the time-resolved average velocity of the plasma between the wires and the probe over the transit time of the fluid parcel, as shown in Figure \ref{fig:velocity}b.

Our simulation shows that the velocity field increases spatially, but decreases temporally. From the simulated velocity field, we estimate $m \approx 2 \times 10^7 \,\SI{}{\per \second}$ and $\nu \approx -4.3 \times 10^6 \,\SI{}{\per \second}$. The inductive probe measurements show that the mean transit time is $\tau \sim \SI{78}{\nano \second}$. This corresponds to a correction factor $R \sim 0.67$ (Equation \ref{eq:ratio}), which indicates that the average velocity measured by the inductive probe is about $67 \%$ of the flow velocity at the probe location. The corrected velocity (green curve in Figure \ref{fig:velocity}b), which estimates the Eulerian velocity at the probe location, ranges from $\sim \SI{170}{\kilo \metre \per \second}$ early in time, to $\sim \SI{78}{\kilo \metre \per \second}$ later in time ($t \sim \SI{300}{\nano \second}$).

\section{\label{sec:imaging} Upstream Mach Number \& Ion Sound Speed}

Figures \ref{fig:interferometry}a and \ref{fig:interferometry}b show the interferograms recorded before the start of the experiment, and at t = 300 ns after start of the current pulse, respectively. The line-integrated electron density, reconstructed from the distorted fringes, is shown in Figure \ref{fig:interferometry}c. A detached bow shock, which appears as a curved discontinuity in the fringes and the electron density, is visible in both the raw interferogram and the electron density map. The shock front appears washed-out in the electron density map because line-integrating through the bulk of the plasma obscures the intensity of the shock front. To enhance the visibility of the shock, we plot the gradient of the line-integrated electron density on a logarithmic scale (Figure \ref{fig:interferometry}d).

\begin{table*}\centering
\ra{1.3}
\caption{\centering Experimentally measured quantities at probe location 300 ns after current start, and comparison with simulation and literature}
\begin{tabular}{rccccc}
\hline
 & Velocity ($\SI{}{\kilo \metre \per \second}$) & Mach Angle (\SI{}{\degree})  & $M_S$ & $C_S$ ($\SI{}{\kilo \metre \per \second}$) & $\bar{Z}T_e$ (eV) \\
 \hline
Experiment  & $52 \pm 14$ & $7^{\circ} \pm 0.5^{\circ}$ & $8.2 \pm 0.6$ & $6 \pm 2$ & $10\pm 6$ \\
Experiment (Corrected)  & $78 \pm 14$ & $7^{\circ} \pm 0.5^{\circ}$ & $8.2 \pm 0.6$ & $9.5 \pm 2$ & $22\pm 9$ \\
Simulation & $68$ & $7^{\circ}$ & $7-11$ & $6-10$ & $9-25$ \\
OTS (Literature) & $61\pm1$ & n/a & n/a & n/a & $28-40$ \\
\hline
\end{tabular}
\label{tab:table}
\end{table*}

The shock front appears more prominent on the top of the probe in Figures \ref{fig:interferometry}b and \ref{fig:interferometry}c. This is because in the shot interferogram, the fringes shift in such a way that they become almost parallel to the shock front under the probe. Consequently, they appear relatively undistorted by the shock. The shock front appears distinct on the other side of the probe, where the fringes are at an angle to the shock. This issue can be eliminated by using a different imaging diagnostic, such as shadowgraphy or schlieren imaging with a symmetric stop.

The shock angle $\sigma(x)$ in a bow shock is the angle between the upstream velocity vector ${\bf u_1}$ and the shock front (Figure \ref{fig:diagnostic_schematic}). The shock angle decreases continuously from $\SI{90}{\degree}$ at the leading edge of the obstacle in a bow shock, and far away from the obstacle, where the shock is infinitesimally weak, the shock angle asymptotically approaches the Mach angle $\mu$. We can relate the Mach angle to the upstream Mach number using the simple expression $\sin \mu = 1/M_S$. \cite{Spreiter1969,Kundu-2012} 
From Figure \ref{fig:interferometry}, we measure the Mach angle to be $\mu = 7.0 \pm \SI{0.5}{\degree}$, which corresponds to an upstream sonic Mach number of $M_S \sim 8.2 \pm 0.6$. Combining this information with the measured flow velocity, we estimate the ion sound speed $C_S$, as shown in Table \ref{tab:table}.

The ion sound speed $C_S$ in a plasma is a function of the electron temperature $T_e$, the ionization $\bar{Z}$, and the adiabatic index $\gamma$ of the plasma. If the value of the adiabatic index $\gamma$ is known, then we can use the estimated sound speed to infer the value of $\bar{Z}T_e$ in the plasma. In high-Z HED plasmas, internal energy and thermal pressure contributions due to Coulomb interactions, ionization, and excitation processes can make the adiabatic index of the plasma smaller than that of an ideal gas. \cite{Drake2006} For the plasma in this experiment ($n_e \sim 1 \times 10^{18} \SI{}{\per \cubic \centi \metre}, T_e \sim 10$ eV), we use a simple model to estimate the value of the adiabatic index $\gamma \approx 1.13$, \cite{Drake2006,Swadling2013} using which, we estimate $\bar{Z}T_e \sim 22 \pm 9$ eV (Table \ref{tab:table}).

\section{\label{sec:sims_lit} Comparison with Simulation \& Literature}

 The velocity measured by the probe (blue curve in Figure \ref{fig:velocity}b) underestimates the simulated Eulerian velocity at the probe location, but applying a correction factor of $R \sim 0.67$ results in good agreement with the simulation. As seen in Table \ref{tab:table}, experimental estimates of the Mach angle, Mach number, and sound speed, also agree well with simulated values. 

The first velocity data point in Figure \ref{fig:velocity}b, however, represents a discrepancy between the experiment and simulation. In the experiment, the plasma reaches the probe earlier than in the simulation. Wire arrays typically exhibit a $\sim \SI{50}{\nano \second}$ dwell time before ablation begins, which means that it is unlikely that there is plasma at the probe location at $t = \SI{50}{\nano \second}$ after start of the current pulse. A magnetostatic Biot-Savart calculation of the vacuum magnetic field shows that the contribution of the vacuum field to the probe's voltage response is only $\sim \SI{20}{\milli \volt}$ at peak current, which is not large enough to explain the rising voltage at $\SI{50}{\nano \second}$. Further investigation is required to resolve the discrepancy in the plasma arrival time between the simulation and experiment.

In Figure \ref{fig:velocity}b, we also compare the flow velocity determined from inductive probe measurements with velocity measurements reported in literature. These measurements are acquired using optical Thompson scattering (OTS) in experiments with aluminum exploding wire arrays of similar dimensions, also fielded on MAGPIE, and at roughly $\SI{5}{\milli \metre}$ from the wires. \cite{Lebedev2014,Burdiak2017,Russell2022} OTS measures spatially-localized flow velocity from the Doppler shift in the spectra of light scattered by the plasma.\cite{Froula2006} The corrected flow velocity estimated from inductive probe measurements agrees, within experimental uncertainty, with OTS measurements reported in literature. \cite{Lebedev2014,Burdiak2017,Russell2022} The ion feature of OTS also allows for the measurement of $\bar{Z}T_e$ from the wavelength separation of the ion acoustic peaks in the scattered spectrum.\cite{Froula2006}  Values of $\bar{Z}T_e$ inferred from OTS in similar plasmas lie in the range $\bar{Z}T_e \sim 28 - 40$ eV, \cite{Burdiak2017} which includes the experimental estimate of $\bar{Z}T_e \sim 22 \pm 9$ eV. 

\section{\label{sec:Summary} Summary}

We present a new experimental technique to infer the velocity and ion sound speed in supersonic magnetized pulsed-power-driven plasmas from simultaneous imaging and voltage measurements of inductive probes. By measuring the transit time of the plasma between two inductive probes, we calculate the average velocity of a magnetized fluid parcel along its trajectory. We show that this average velocity can be related to the Eulerian velocity field, and the extent to which the measured velocity over- or underestimates the instantaneous flow velocity at the probe location depends on the temporal and spatial variation of the velocity field. In this demonstration, we use a Mach-Zehnder interferometer to image the shock --- however, other imaging techniques such as shadowgraphy or schlieren imaging, where the intensity of the image is proportional to density gradients (such as at the shock front), may enhance the visibility of the shock. We estimate the Mach number of the upstream flow from the Mach angle of the imaged shock, and combining this information with the measured velocity allows us to estimate the ion sound speed, which agrees well with the prediction of our simulation, and with OTS measurements reported in literature. This diagnostic technique is especially useful in experiments where an OTS diagnostic is either unavailable or unfeasible. In this paper, we have verified the use of inductive probes to measure time-resolved velocity and ion sound speed on a $\sim 1$ MA university-scale pulsed-power machine. In future experiments, we plan to utilize this technique for the MARZ experimental campaign \cite{hare2021simulations} on the $\sim 30$ MA Z pulsed-power facility, to measure both space- and time-resolved velocity and ion sound speed using an array of inductive probes fielded at different radial locations around exploding wire arrays. 

\section{Acknowledgements}

This work was funded in part by NSF and NNSA under grant no. PHY2108050, and supported by the U.S. Department of Energy (DOE) under Award Nos. DE-SC0020434, DE-NA0003764, DE-F03-02NA00057, DE-SC-0001063, and DE-NA0003868, and the Engineering and Physical Sciences Research Council (EPSRC) under Grant No. EP/N013379/1. The simulations presented in this paper were performed on the MIT-PSFC partition of the Engaging cluster at the MGHPCC facility (www.mghpcc.org) which was funded by DOE grant no. DE-FG02-91-ER54109. The authors would also like to thank I. Tang for assistance with the experiments.

\bibliography{aipsamp}

%merlin.mbs aipnum4-1.bst 2010-07-25 4.21a (PWD, AO, DPC) hacked
%Control: key (0)
%Control: author (8) initials jnrlst
%Control: editor formatted (1) identically to author
%Control: production of article title (0) allowed
%Control: page (1) range
%Control: year (1) truncated
%Control: production of eprint (0) enabled
\begin{thebibliography}{30}%
\makeatletter
\providecommand \@ifxundefined [1]{%
 \@ifx{#1\undefined}
}%
\providecommand \@ifnum [1]{%
 \ifnum #1\expandafter \@firstoftwo
 \else \expandafter \@secondoftwo
 \fi
}%
\providecommand \@ifx [1]{%
 \ifx #1\expandafter \@firstoftwo
 \else \expandafter \@secondoftwo
 \fi
}%
\providecommand \natexlab [1]{#1}%
\providecommand \enquote  [1]{``#1''}%
\providecommand \bibnamefont  [1]{#1}%
\providecommand \bibfnamefont [1]{#1}%
\providecommand \citenamefont [1]{#1}%
\providecommand \href@noop [0]{\@secondoftwo}%
\providecommand \href [0]{\begingroup \@sanitize@url \@href}%
\providecommand \@href[1]{\@@startlink{#1}\@@href}%
\providecommand \@@href[1]{\endgroup#1\@@endlink}%
\providecommand \@sanitize@url [0]{\catcode `\\12\catcode `\$12\catcode
  `\&12\catcode `\#12\catcode `\^12\catcode `\_12\catcode `\%12\relax}%
\providecommand \@@startlink[1]{}%
\providecommand \@@endlink[0]{}%
\providecommand \url  [0]{\begingroup\@sanitize@url \@url }%
\providecommand \@url [1]{\endgroup\@href {#1}{\urlprefix }}%
\providecommand \urlprefix  [0]{URL }%
\providecommand \Eprint [0]{\href }%
\providecommand \doibase [0]{http://dx.doi.org/}%
\providecommand \selectlanguage [0]{\@gobble}%
\providecommand \bibinfo  [0]{\@secondoftwo}%
\providecommand \bibfield  [0]{\@secondoftwo}%
\providecommand \translation [1]{[#1]}%
\providecommand \BibitemOpen [0]{}%
\providecommand \bibitemStop [0]{}%
\providecommand \bibitemNoStop [0]{.\EOS\space}%
\providecommand \EOS [0]{\spacefactor3000\relax}%
\providecommand \BibitemShut  [1]{\csname bibitem#1\endcsname}%
\let\auto@bib@innerbib\@empty
%</preamble>
\bibitem [{\citenamefont {Hare}\ \emph {et~al.}(2018)\citenamefont {Hare} \emph
  {et~al.}}]{Hare2018}%
  \BibitemOpen
  \bibfield  {author} {\bibinfo {author} {\bibfnamefont {J.~D.}\ \bibnamefont
  {Hare}} \emph {et~al.},\ }\href {http://dx.doi.org/10.1063/1.5016280}
  {\bibfield  {journal} {\bibinfo  {journal} {Phys. Plasmas}\ }\textbf
  {\bibinfo {volume} {25}} (\bibinfo {year} {2018})}\BibitemShut {NoStop}%
\bibitem [{\citenamefont {Kugland}\ \emph {et~al.}(2012)\citenamefont {Kugland}
  \emph {et~al.}}]{Kugland2012}%
  \BibitemOpen
  \bibfield  {author} {\bibinfo {author} {\bibfnamefont {N.~L.}\ \bibnamefont
  {Kugland}} \emph {et~al.},\ }\href {http://dx.doi.org/10.1063/1.4750234}
  {\bibfield  {journal} {\bibinfo  {journal} {Rev. Sci. Instrum.}\ }\textbf
  {\bibinfo {volume} {83}} (\bibinfo {year} {2012})}\BibitemShut {NoStop}%
\bibitem [{\citenamefont {Sutcliffe}\ \emph {et~al.}(2021)\citenamefont
  {Sutcliffe} \emph {et~al.}}]{Sutcliffe2021}%
  \BibitemOpen
  \bibfield  {author} {\bibinfo {author} {\bibfnamefont {G.}~\bibnamefont
  {Sutcliffe}} \emph {et~al.},\ }\href {https://doi.org/10.1063/5.0043845}
  {\bibfield  {journal} {\bibinfo  {journal} {Rev. Sci. Instrum.}\ }\textbf
  {\bibinfo {volume} {92}} (\bibinfo {year} {2021})}\BibitemShut {NoStop}%
\bibitem [{\citenamefont {Everson}\ \emph {et~al.}(2009)\citenamefont {Everson}
  \emph {et~al.}}]{Everson2009}%
  \BibitemOpen
  \bibfield  {author} {\bibinfo {author} {\bibfnamefont {E.~T.}\ \bibnamefont
  {Everson}} \emph {et~al.},\ }\href {http://dx.doi.org/10.1063/1.3246785}
  {\bibfield  {journal} {\bibinfo  {journal} {Rev. Sci. Instrum.}\ }\textbf
  {\bibinfo {volume} {80}} (\bibinfo {year} {2009})}\BibitemShut {NoStop}%
\bibitem [{\citenamefont {Pilgram}\ \emph {et~al.}(2021)\citenamefont {Pilgram}
  \emph {et~al.}}]{Pilgram2021}%
  \BibitemOpen
  \bibfield  {author} {\bibinfo {author} {\bibfnamefont {J.~J.}\ \bibnamefont
  {Pilgram}} \emph {et~al.},\ }\href {http://arxiv.org/abs/2111.02670} {\
  (\bibinfo {year} {2021})},\ \Eprint {http://arxiv.org/abs/2111.02670}
  {arXiv:2111.02670} \BibitemShut {NoStop}%
\bibitem [{\citenamefont {Levesque}\ \emph {et~al.}(2022)\citenamefont
  {Levesque} \emph {et~al.}}]{Levesque2022}%
  \BibitemOpen
  \bibfield  {author} {\bibinfo {author} {\bibfnamefont {J.~M.}\ \bibnamefont
  {Levesque}} \emph {et~al.},\ }\href {http://dx.doi.org/10.1063/5.0062254}
  {\bibfield  {journal} {\bibinfo  {journal} {Phys. Plasmas}\ }\textbf
  {\bibinfo {volume} {29}} (\bibinfo {year} {2022})}\BibitemShut {NoStop}%
\bibitem [{\citenamefont {Schaeffer}\ \emph {et~al.}(2022)\citenamefont
  {Schaeffer} \emph {et~al.}}]{Schaeffer2022}%
  \BibitemOpen
  \bibfield  {author} {\bibinfo {author} {\bibfnamefont {D.~B.}\ \bibnamefont
  {Schaeffer}} \emph {et~al.},\ }\href {http://arxiv.org/abs/2201.02176} {\
  (\bibinfo {year} {2022})},\ \Eprint {http://arxiv.org/abs/2201.02176}
  {arXiv:2201.02176} \BibitemShut {NoStop}%
\bibitem [{\citenamefont {Burdiak}\ \emph {et~al.}(2017)\citenamefont {Burdiak}
  \emph {et~al.}}]{Burdiak2017}%
  \BibitemOpen
  \bibfield  {author} {\bibinfo {author} {\bibfnamefont {G.~C.}\ \bibnamefont
  {Burdiak}} \emph {et~al.},\ }\href {http://dx.doi.org/10.1063/1.4993187}
  {\bibfield  {journal} {\bibinfo  {journal} {Phys. Plasmas}\ }\textbf
  {\bibinfo {volume} {24}} (\bibinfo {year} {2017})}\BibitemShut {NoStop}%
\bibitem [{\citenamefont {Suttle}\ \emph {et~al.}(2019)\citenamefont {Suttle}
  \emph {et~al.}}]{Suttle_2019}%
  \BibitemOpen
  \bibfield  {author} {\bibinfo {author} {\bibfnamefont {L.~G.}\ \bibnamefont
  {Suttle}} \emph {et~al.},\ }\href {https://doi.org/10.1088/1361-6587/ab5296}
  {\bibfield  {journal} {\bibinfo  {journal} {Plasma Phys. Control. Fusion}\
  }\textbf {\bibinfo {volume} {62}} (\bibinfo {year} {2019})}\BibitemShut
  {NoStop}%
\bibitem [{\citenamefont {Ji}\ \emph {et~al.}(1999)\citenamefont {Ji} \emph
  {et~al.}}]{Ji1999}%
  \BibitemOpen
  \bibfield  {author} {\bibinfo {author} {\bibfnamefont {H.}~\bibnamefont {Ji}}
  \emph {et~al.},\ }\href {http://dx.doi.org/10.1063/1.873432} {\bibfield
  {journal} {\bibinfo  {journal} {Phys. Plasmas}\ }\textbf {\bibinfo {volume}
  {6}} (\bibinfo {year} {1999})}\BibitemShut {NoStop}%
\bibitem [{\citenamefont {Froula}\ \emph {et~al.}(2006)\citenamefont {Froula}
  \emph {et~al.}}]{Froula2006}%
  \BibitemOpen
  \bibfield  {author} {\bibinfo {author} {\bibfnamefont {D.~H.}\ \bibnamefont
  {Froula}} \emph {et~al.},\ }\href {http://dx.doi.org/10.1063/1.2336451}
  {\bibfield  {journal} {\bibinfo  {journal} {Rev. Sci. Instrum.}\ }\textbf
  {\bibinfo {volume} {77}} (\bibinfo {year} {2006})}\BibitemShut {NoStop}%
\bibitem [{\citenamefont {Rosenberg}\ \emph {et~al.}(2012)\citenamefont
  {Rosenberg} \emph {et~al.}}]{Rosenberg2012}%
  \BibitemOpen
  \bibfield  {author} {\bibinfo {author} {\bibfnamefont {M.~J.}\ \bibnamefont
  {Rosenberg}} \emph {et~al.},\ }\href
  {http://dx.doi.org/10.1103/PhysRevE.86.056407} {\bibfield  {journal}
  {\bibinfo  {journal} {Phys. Rev. E}\ }\textbf {\bibinfo {volume} {86}}
  (\bibinfo {year} {2012})}\BibitemShut {NoStop}%
\bibitem [{\citenamefont {Suttle}\ \emph {et~al.}(2021)\citenamefont {Suttle}
  \emph {et~al.}}]{Suttle2021}%
  \BibitemOpen
  \bibfield  {author} {\bibinfo {author} {\bibfnamefont {L.~G.}\ \bibnamefont
  {Suttle}} \emph {et~al.},\ }\href {https://doi.org/10.1063/5.0041118}
  {\bibfield  {journal} {\bibinfo  {journal} {Rev. Sci. Instrum.}\ }\textbf
  {\bibinfo {volume} {92}} (\bibinfo {year} {2021})}\BibitemShut {NoStop}%
\bibitem [{\citenamefont {Hutchinson}(2002)}]{Hutchinson2002}%
  \BibitemOpen
  \bibfield  {author} {\bibinfo {author} {\bibfnamefont {I.~H.}\ \bibnamefont
  {Hutchinson}},\ }\href {https://doi.org/10.1017/CBO9780511613630} {\emph
  {\bibinfo {title} {Principles of Plasma Diagnostics}}}\ (\bibinfo {year}
  {2002})\BibitemShut {NoStop}%
\bibitem [{\citenamefont {Suckewer}\ \emph {et~al.}(1979)\citenamefont
  {Suckewer} \emph {et~al.}}]{Suckewer1979}%
  \BibitemOpen
  \bibfield  {author} {\bibinfo {author} {\bibfnamefont {S.}~\bibnamefont
  {Suckewer}} \emph {et~al.},\ }\href
  {http://dx.doi.org/10.1103/PhysRevLett.43.207} {\bibfield  {journal}
  {\bibinfo  {journal} {Phys. Rev. Lett.}\ }\textbf {\bibinfo {volume} {43}}
  (\bibinfo {year} {1979})}\BibitemShut {NoStop}%
\bibitem [{\citenamefont {Yamada}\ \emph {et~al.}(1997)\citenamefont {Yamada}
  \emph {et~al.}}]{Yamada1997}%
  \BibitemOpen
  \bibfield  {author} {\bibinfo {author} {\bibfnamefont {M.}~\bibnamefont
  {Yamada}} \emph {et~al.},\ }\href {http://dx.doi.org/10.1063/1.872336}
  {\bibfield  {journal} {\bibinfo  {journal} {Phys. Plasmas}\ }\textbf
  {\bibinfo {volume} {4}} (\bibinfo {year} {1997})}\BibitemShut {NoStop}%
\bibitem [{\citenamefont {Goedbloed}\ \emph {et~al.}(2010)\citenamefont
  {Goedbloed} \emph {et~al.}}]{goedbloed_keppens_poedts_2010}%
  \BibitemOpen
  \bibfield  {author} {\bibinfo {author} {\bibfnamefont {J.~P.}\ \bibnamefont
  {Goedbloed}} \emph {et~al.},\ }\href {\doibase 10.1017/CBO9781139195560}
  {\emph {\bibinfo {title} {Advanced Magnetohydrodynamics}}}\ (\bibinfo {year}
  {2010})\BibitemShut {NoStop}%
\bibitem [{\citenamefont {Spreiter}\ and\ \citenamefont
  {Alksne}(1969)}]{Spreiter1969}%
  \BibitemOpen
  \bibfield  {author} {\bibinfo {author} {\bibfnamefont {J.~R.}\ \bibnamefont
  {Spreiter}}\ and\ \bibinfo {author} {\bibfnamefont {A.~Y.}\ \bibnamefont
  {Alksne}},\ }\href {http://dx.doi.org/10.1029/RG007i001p00011} {\bibfield
  {journal} {\bibinfo  {journal} {Rev. Geophys.}\ }\textbf {\bibinfo {volume}
  {7}} (\bibinfo {year} {1969})}\BibitemShut {NoStop}%
\bibitem [{\citenamefont {Anderson}(2001)}]{anderson_2001}%
  \BibitemOpen
  \bibfield  {author} {\bibinfo {author} {\bibfnamefont {J.~D.}\ \bibnamefont
  {Anderson}},\ }\href {http://www.worldcat.org/isbn/9780073398105} {\emph
  {\bibinfo {title} {Fundamentals of Aerodynamics}}}\ (\bibinfo {year}
  {2001})\BibitemShut {NoStop}%
\bibitem [{\citenamefont {Lebedev}\ \emph {et~al.}(2014)\citenamefont {Lebedev}
  \emph {et~al.}}]{Lebedev2014}%
  \BibitemOpen
  \bibfield  {author} {\bibinfo {author} {\bibfnamefont {S.~V.}\ \bibnamefont
  {Lebedev}} \emph {et~al.},\ }\href {http://dx.doi.org/10.1063/1.4874334}
  {\bibfield  {journal} {\bibinfo  {journal} {Phys. Plasmas}\ }\textbf
  {\bibinfo {volume} {21}} (\bibinfo {year} {2014})}\BibitemShut {NoStop}%
\bibitem [{\citenamefont {Russell}\ \emph {et~al.}(2022)\citenamefont {Russell}
  \emph {et~al.}}]{Russell2022}%
  \BibitemOpen
  \bibfield  {author} {\bibinfo {author} {\bibfnamefont {D.~R.}\ \bibnamefont
  {Russell}} \emph {et~al.},\ }\href@noop {} {\  (\bibinfo {year} {2022})},\
  \Eprint {http://arxiv.org/abs/arXiv:2201.09039v1} {arXiv:2201.09039v1}
  \BibitemShut {NoStop}%
\bibitem [{\citenamefont {Mitchell}\ \emph {et~al.}(1996)\citenamefont
  {Mitchell} \emph {et~al.}}]{Mitchell1996}%
  \BibitemOpen
  \bibfield  {author} {\bibinfo {author} {\bibfnamefont {I.~H.}\ \bibnamefont
  {Mitchell}} \emph {et~al.},\ }\href {http://dx.doi.org/10.1063/1.1146884}
  {\bibfield  {journal} {\bibinfo  {journal} {Rev. Sci. Instrum.}\ }\textbf
  {\bibinfo {volume} {67}} (\bibinfo {year} {1996})}\BibitemShut {NoStop}%
\bibitem [{\citenamefont {Swadling}\ \emph {et~al.}(2013)\citenamefont
  {Swadling} \emph {et~al.}}]{Swadling2013}%
  \BibitemOpen
  \bibfield  {author} {\bibinfo {author} {\bibfnamefont {G.~F.}\ \bibnamefont
  {Swadling}} \emph {et~al.},\ }\href {http://dx.doi.org/10.1063/1.4790520}
  {\bibfield  {journal} {\bibinfo  {journal} {Phys. Plasmas}\ }\textbf
  {\bibinfo {volume} {20}} (\bibinfo {year} {2013})}\BibitemShut {NoStop}%
\bibitem [{\citenamefont {Hare}\ \emph {et~al.}(2019)\citenamefont {Hare} \emph
  {et~al.}}]{Hare2019}%
  \BibitemOpen
  \bibfield  {author} {\bibinfo {author} {\bibfnamefont {J.~D.}\ \bibnamefont
  {Hare}} \emph {et~al.},\ }\href {https://dx.doi.org/10.1088/1361-6587/ab2571}
  {\bibfield  {journal} {\bibinfo  {journal} {Plasma Phys. Control. Fusion}\
  }\textbf {\bibinfo {volume} {61}} (\bibinfo {year} {2019})}\BibitemShut
  {NoStop}%
\bibitem [{\citenamefont {Dursi}\ and\ \citenamefont
  {Pfrommer}(2008)}]{Dursi2008}%
  \BibitemOpen
  \bibfield  {author} {\bibinfo {author} {\bibfnamefont {L.~J.}\ \bibnamefont
  {Dursi}}\ and\ \bibinfo {author} {\bibfnamefont {C.}~\bibnamefont
  {Pfrommer}},\ }\href {https://doi.org/10.1086/529371} {\bibfield  {journal}
  {\bibinfo  {journal} {Astrophys. J.}\ }\textbf {\bibinfo {volume} {677}}
  (\bibinfo {year} {2008})}\BibitemShut {NoStop}%
\bibitem [{\citenamefont {Chittenden}\ \emph {et~al.}(2004)\citenamefont
  {Chittenden} \emph {et~al.}}]{Chittenden2004}%
  \BibitemOpen
  \bibfield  {author} {\bibinfo {author} {\bibfnamefont {J.~P.}\ \bibnamefont
  {Chittenden}} \emph {et~al.},\ }\href {https://doi.org/10.1063/1.1643756}
  {\bibfield  {journal} {\bibinfo  {journal} {Phys. Plasmas}\ }\textbf
  {\bibinfo {volume} {11}} (\bibinfo {year} {2004})}\BibitemShut {NoStop}%
\bibitem [{\citenamefont {Huba}(2013)}]{Huba2013}%
  \BibitemOpen
  \bibfield  {author} {\bibinfo {author} {\bibfnamefont {J.}~\bibnamefont
  {Huba}},\ }\href {https://doi.org/10.1109/mei.2003.1178121} {\bibfield
  {journal} {\bibinfo  {journal} {IEEE Electrical Insulation Magazine}\ }
  (\bibinfo {year} {2013})}\BibitemShut {NoStop}%
\bibitem [{\citenamefont {Kundu}\ \emph {et~al.}(2012)\citenamefont {Kundu}
  \emph {et~al.}}]{Kundu-2012}%
  \BibitemOpen
  \bibfield  {author} {\bibinfo {author} {\bibfnamefont {P.~K.}\ \bibnamefont
  {Kundu}} \emph {et~al.},\ }\href
  {https://doi.org/10.1016/B978-0-12-382100-3.10015-0} {\emph {\bibinfo {title}
  {Fluid Mechanics (Fifth Edition)}}}\ (\bibinfo {year} {2012})\BibitemShut
  {NoStop}%
\bibitem [{\citenamefont {Drake}(2013)}]{Drake2006}%
  \BibitemOpen
  \bibfield  {author} {\bibinfo {author} {\bibfnamefont {R.}~\bibnamefont
  {Drake}},\ }\href {https://doi.org/10.1007/978-3-319-67711-8} {\emph
  {\bibinfo {title} {{High-Energy-Density Physics}}}}\ (\bibinfo {year}
  {2013})\BibitemShut {NoStop}%
\bibitem [{\citenamefont {Hare}\ \emph {et~al.}(2021)\citenamefont {Hare} \emph
  {et~al.}}]{hare2021simulations}%
  \BibitemOpen
  \bibfield  {author} {\bibinfo {author} {\bibfnamefont {J.~D.}\ \bibnamefont
  {Hare}} \emph {et~al.},\ }\href
  {https://meetings.aps.org/Meeting/DPP21/Session/NP11.103} {\bibfield
  {journal} {\bibinfo  {journal} {Bulletin of the American Physical Society}\
  }\textbf {\bibinfo {volume} {66}} (\bibinfo {year} {2021})}\BibitemShut
  {NoStop}%
\end{thebibliography}%

\end{document}